\documentclass[11pt,a4paper]{article} 
\pdfoutput=1
\usepackage[utf8x]{inputenc}     % if unicode errors are coming
\usepackage{jheppub}
\usepackage{enumerate}
\usepackage{epsfig} 
\usepackage{float} 
\usepackage[caption = false]{subfig}
\usepackage{wasysym} 
\usepackage{mathrsfs} 
\usepackage{amsfonts} 
\usepackage{amsbsy} 
\usepackage{amscd} 
\usepackage{pstricks} 
\usepackage{multirow} 
\usepackage{tikz}
\usepackage{color}
\usepackage{slashed}
\usepackage{array}
\usepackage{tablefootnote}

\usetikzlibrary{arrows,positioning,shapes.geometric} 
\usepackage[compat=1.1.0]{tikz-feynman}          % package allowing Feynman diagrams
\usepackage[font=small,labelfont=bf]{caption}
\usepackage{slashed}
\usepackage{multirow}
\usepackage{tabularx}
\usepackage{soul}  % Strikethrough text with \st{Hellow world}
\usepackage{listings} 
\usepackage{color}
\title{Energy-weighted Message Passing: an infra-red and collinear safe graph neural network algorithm}

\author[a]{Partha Konar,}
\author[a,b]{Vishal~S.~Ngairangbam,}
\author[c,d]{and Michael~Spannowsky} 

\affiliation[a]{Theoretical Physics Division, Physical Research Laboratory,\\ Shree Pannalal Patel Marg, Ahmedabad - 380009, Gujarat, India}
\affiliation[b]{Discipline of Physics, Indian Institute of Technology, Palaj,\\ Gandhinagar - 382424, Gujarat, India}
\affiliation[c]{Institute for Particle Physics Phenomenology, Durham University,\\ Durham DH1 3LE, United Kingdom}
\affiliation[d]{Department of Physics, Durham University,\\ Durham DH1 3LE, United Kingdom}
\emailAdd{konar@prl.res.in}
\emailAdd{vishalng@prl.res.in}
\emailAdd{michael.spannowsky@durham.ac.uk}

\abstract{Hadronic signals of new-physics origin at the Large Hadron Collider can remain hidden within the copiously produced hadronic jets. Unveiling such signatures require highly performant deep-learning algorithms. We construct a class of Graph Neural Networks (GNN) in the message-passing formalism that makes the network output infra-red and collinear (IRC) safe, an important criterion satisfied within perturbative QCD calculations. Including IRC safety of the network output as a requirement in the construction of the GNN improves its explainability and robustness against theoretical uncertainties in the data.
We generalise Energy Flow Networks (EFN), an IRC safe deep-learning algorithm on a point cloud, defining energy weighted local and global readouts on GNNs. Applying the simplest of such networks to identify top quarks, W bosons and quark/gluon jets, we find that it outperforms state-of-the-art EFNs. Additionally, we obtain a general class of graph construction algorithms that give structurally invariant graphs in the IRC limit, a necessary criterion for the IRC safety of the GNN output.}

\preprint{IPPP/21/33}

\keywords{ Large Hadron Collider, Hadronic jets, Message-passing Graph Neural Networks}
\allowdisplaybreaks

\begin{document}\maketitle

\flushbottom
\section{Introduction}

The Large Hadron Collider (LHC) has been collecting an immense amount of data for the past decade and will continue to do so for the upcoming decade. With the absence of signatures of new physics (NP), it is imperative to critically analyse all the available data and reduce the possibility of missing out by exploring all possible ways of analysis. Machine learning (ML) algorithms are powerful statistical learning tools that can extract features directly from data via an optimisation procedure and are used to efficiently analyse the huge amounts of collected data~\cite{Guest:2018yhq,ATLAS:2019vwv,ATLAS:2020pcy,ATLAS:2020iwa}. There has been a growing focus~\cite{deOliveira:2015xxd,Komiske:2018vkc,Ngairangbam:2020ksz,Diefenbacher:2019ezd,Blance:2019ibf,Kasieczka:2017nvn,ATL-PHYS-PUB-2020-014,Louppe:2017ipp,Araz:2021wqm} on adapting the more powerful versions of ML\footnote{See~Ref.~\cite{Feickert:2021ajf} for an updated list of application of machine learning in particle physics.}: the so-called deep-learning algorithms that extract features from low-level data like the four-momenta of detected particles. Although highly performant, deep-learning algorithms are not physically well-understood due to many tunable parameters and the numeric nature of the optimisation process. Thus, one of the biggest challenges in deploying machine-learning methods to searches for new physics and measurements of particle properties is their inherent complexity and the accompanying valuation to be a so-called `black box'. Consequently, to facilitate the application of neural network methods reliably to the analysis of data at particle colliders, there is an ongoing focus to understand the phenomenologically relevant characteristics of neural networks, i.e. interpretability~\cite{Butter:2017cot,Erdmann:2018shi,Chakraborty:2019imr,Faucett:2020vbu,Lai:2020byl,Kasieczka:2020nyd}, robustness to soft and collinear emissions~\cite{Choi:2018dag,Komiske:2018cqr,Larkoski:2019nwj,Dolan:2020qkr,Romero:2021qlf,Shimmin:2021pkm}, and better control of uncertainties~\cite{Englert:2018cfo,Blance:2019ibf, Bollweg:2019skg,Ghosh:2021roe,Nachman:2019dol}. Extending these efforts, we devise Graph Neural Networks, the current state-of-the-art deep-learning algorithm in various applications at LHC, providing Neural Network outputs robust to soft and collinear emissions.

At the LHC, evidence for new physics can be hidden in the dominantly produced jets, collimated sprays of hadrons arising from energetic, strongly interacting particles. Considerable attention to the characteristics of jets is required for a holistic search strategy. Jet-substructure~\cite{Butterworth:2008iy,Thaler:2010tr,Plehn:2011sj,Marzani:2019hun} is a sub-area of LHC physics that has received significant investigation both from a theory-driven perspective~\cite{Dasgupta:2013ihk,Larkoski:2014gra,Larkoski:2014wba,Larkoski:2015lea,Dasgupta:2013via,Dasgupta:2021kgi,Sterman:2021gyk} as well as the data-driven paradigm~\cite{Baldi:2016fql,Larkoski:2017jix,Komiske:2016rsd,Barnard:2016qma,Macaluso:2018tck,Shimmin:2017mfk,Datta:2017rhs,Khosa:2021cyk}. It focuses on distinguishing rare hadronic decays of heavy boosted particles (signal) like the top-quark and Higgs boson from those originating from the abundantly produced (background) gluons and lighter quarks. Conventional analyses~\cite{Plehn:2009rk,Das:2017gke,Bhardwaj:2018lma,Bhardwaj:2020llc,Adams:2015hiv} use variables that distinguish between the evolution of emissions of the signal and background jets. These high-level variables are calculated from the four-momenta of the constituents and are amenable to calculations~\cite{Dasgupta:2013ihk} in perturbative QCD. Deep-learning algorithms have been found to outperform the high-level variables at the cost of losing the intuitive physics understanding of the learnt features. Thus, in general, we do not control the possibility of the feature extraction process, which leaves our classification performance potentially highly susceptible to non-perturbative effects. A necessary condition for any observable to be calculable within perturbative QCD is infra-red and collinear (IRC) safety. It ensures the appropriate handling of real and virtual corrections order by order in perturbative calculations via the KLN theorem~\cite{doi:10.1063/1.1724268,PhysRev.133.B1549}. Thus, an IRC safe deep-learning algorithm would learn features that are, in principle, calculable in perturbative QCD.

The most comprehensive representation of collision events at the LHC are sets of individually detected particles or physically reconstructed objects. The number of elements in these sets vary from event to event and are denoted naturally in terms of \emph{point clouds} with each entry containing a vector composed of observables like four-momenta, charge, and the identity of the object. Graphs are discrete mathematical data structures defined on these sets' elements (nodes), where the edges encode relational information. When constructing graphs from a point cloud, its primary role is to unveil \emph{local structures} within the data. A popular algorithm for constructing a graph from a point cloud also used to construct jet graphs for classification by Graph Neural Networks (GNNs)~\cite{Qu:2019gqs,Bernreuther:2020vhm}, ``the k-nearest neighbour graph", has a fixed number of connections by definition and hence the GNN's output will not be IRC safe. The GNN output's IRC safety is closely connected to the graph construction algorithm. This has not been explored previously, to our knowledge, and we find that only specific classes of graph construction algorithms guarantee the IRC safety of the network output. 
More accurately, the graph determines how the subnetworks are applied to the data via the edge connections. If the edges change drastically in the presence of soft or collinear particles, the network output will also change. The graph construction is analogous to the jet algorithm, while the output is analogous to observables defined on the jet. 

GNNs~\cite{gnn_first,gilmer2017neural,Shlomi:2020gdn} have been studied for jet classification in supervised~\cite{henrion2017neural,Qu:2019gqs,Bernreuther:2020vhm,Dreyer:2020brq,Moreno:2019bmu,Moreno:2019neq} as well as unsupervised~\cite{Atkinson:2021nlt,Blance:2021gcs} scenarios and has state-of-the-art performance~\cite{Qu:2019gqs} compared to other still excellent architectures~\cite{Kasieczka:2019dbj} like Convolutional Neural Networks (CNNs), Deep-sets, and Recurrent Neural Networks (RNNs). The better performance originates in GNNs having an inductive bias more appropriate for jet substructure in particular and collider physics in general. These biases include generalising the Euclidean bias of CNNs to higher dimensional non-Euclidean spaces~\cite{geom_deep}, enhancing the feature extraction in the deep-sets~\cite{charles2017pointnet,zaheer2017deep,qi2017pointnet++} framework by including local structures~\cite{wang2019dynamic}, and generalising RNNs~\cite{572108,712151} to undirected cyclic graphs~\cite{gnn_first}. Hence, incorporating IRC safety into GNNs will further strengthen its biases for various applications at LHC. GNNs are a special type of hierarchical neural networks \cite{Mavrovouniotis1992}, consisting of several subnetworks organised so that the output respects properties of graph-structured data like permutation-invariance of the nodes. It has received wide attention in recent years and has been concisely described in the message-passing neural network (MPNN) formalism~\cite{gilmer2017neural}. Thus, we will examine the MPNN formalism concentrating on the problem of inductive\footnote{One significant area which uses GNNs is in transductive learning~\cite{kipf2017semisupervised}, where the aim is to extend information to partially known regions of data. Such networks perform poorly on unseen data, as it learns features specific to the data. For most purposes in collider phenomenology, we are always in an inductive learning domain, where the model is applied to unseen data.} graph classification. 

A closely connected algorithm to GNNs: the deep-sets framework for feature learning on point clouds, has been explored for jet physics. \emph{Energy Flow Networks} (EFNs)~\cite{Komiske:2018cqr} are IRC safe deep learning models for point clouds, where the feature extraction component learns a per-particle-map to a latent space. The process of constructing graphs out of the point cloud imposes additional structures into the data, which can be efficiently extracted with the help of MPNNs. Concretely, an MPPN based feature extraction phase improves the per-particle-map in the following ways:
\begin{itemize}
	\item It can extract inter particle information courtesy of the trainable message-passing function $\Phi(p_i,p_j)$, acting on each pair of nodes $p_i$ and $p_j$ connected by each of the edges in the graph. 
	\item The node readout updates the node feature as a permutation invariant function of all incoming messages. The readout, along with the message-passing step, forms one \emph{message-passing} operation. It controls the extent of information passed from one layer to another. Therefore, the graph construction algorithm directly controls the nature of the information that goes into learning the parameters of the message function of the first layer.
	\item Since the updated node features are functions of all the neighbouring node features, the range of information in the node features gradually increases with the repetitive application of the message-passing operation. This is not the case for EFNs, as the function is dependent on single-node features. Thus, applying a subsequent learnable function to the updated node features becomes a functional composition, which does not add additional complexity to the process of feature extraction. 
	\item On top of the graph construction itself, the number of applications of the message-passing operation also controls the amount of local information encoded into the final node features. For EFNs, this is always limited to single particles. 	
\end{itemize}  
Forgoing the permutation invariance of EFNs, for permutation equivariance~\cite{Dolan:2020qkr} has better feature extraction by partially taking care of the last two points, at the additional cost of having to abandon the variable-length inputs. On the contrary, an IRC safe MPNN would improve upon the EFNs and still be permutation invariant. Intrinsically, this is because they are very similar, which is also self-evident within the discussed reasons. Once the graph construction algorithm is taken care of, we find that implementing an IRC safe MPNN can be done via an energy-weighted message (feature) with summed aggregation at the node (graph) level. We find that the network, which we refer to as \emph{Energy-weighted Message-Passing Network} (EMPN), improves upon EFNs with a single message-passing operation. Moreover, EMPNs can, in principle, improve upon EFNs in all the four points discussed above as the iterative application does not spoil the IRC safety.  

The rest of the paper is organised as follows. Section \ref{sec:mpnn} introduces the basic idea behind graphs and MPNNs. Readers familiar with graphs and graph neural networks can skip this section. We present the main results of this work in Section \ref{sec:ewpm}, where we devise the graph construction algorithm and the MPNN architecture, which guarantees the IRC safety of the network output. We describe in detail the application of EMPNs to three jet-tagging scenarios: gluon vs quark, QCD vs $W$, and QCD vs top, in Section \ref{sec:net_imp}. The results of these three scenarios are presented in Section \ref{sec:results}. We conclude in Section \ref{sec:conclusion}.

\section{A brief recap of message-passing neural networks}
\label{sec:mpnn}
In this section, we present graphs in the context of particle four-vectors along with a self-contained introduction to message-passing neural networks. As the literature on graphs and graph neural networks is rich and expansive, we focus on the context of collider phenomenology and jet physics. Finally, we give a presentation of graphs and graph neural networks, which is essential for understanding the main result of our work.

\subsection{Graphs}
 A graph $G(\mathcal{S},\mathcal{E})$ is defined on a set of nodes $\mathcal{S}$ with edges $\mathcal{E}\subseteq\mathcal{S}\times\mathcal{S}=\{(i,j)\;|\;i,j\in\mathcal{S}\}$ consisting of an ordered pair of elements in $\mathcal{S}$. The nodes can have a representation $\mathbf{h}_i\in \mathcal{M}\;\forall\;i\in\mathcal{S}$, in some metric space $\mathcal{M}$, where $\mathbf{h}_i$ is the feature of node $i$. In the context of our present analysis, this metric-space is the union of the timelike and lightlike regions\footnote{In general, it could include other information like charge, detector component etc., which we exclude for IRC safety. Although not exactly lightlike, we assume that the jet constituents are massless for easier analysis.} of the \emph{Lorentz manifold} with the Minkowski metric or some other metric in flat spacetime. This could extend in principle to the spacelike regions for theoretically motivated considerations or be generalised to include general coordinate transformations in curved spacetime for applications in the domain of general relativity. When learning from a point cloud, the edge set $\mathcal{E}$ is not provided \emph{a priori} and is constructed with an algorithm defined on the node features. Well-known examples exist in the point cloud literature~\cite{NATALI2011151}, the most famous one being the $k$-nearest neighbour (k-NN) graph~\cite{wang2019dynamic, Qu:2019gqs}. The edges can also have a representation $\mathbf{e}_{ij}$ in some space $\mathcal{X}$. In our context, these can be quantities like mass, directional separation, or the generalised $k_t$ measures~\cite{Cacciari:2008gp} which are derived from the node features themselves. 
 
 A \emph{walk} is a sequence of edges that joins a sequence of nodes, for instance, $$W=((i,j),(j,l),(l,m),(m,i),(i,j))\quad,$$ is a walk with the edge $(i,j)$ repeated twice. On a jet graph, all possible walks of length $L$ would indicate all possible flow of information between the nodes under $L$ iterative application of message-passing operations. If all the edges are distinct, a walk is called a \emph{trail}. A \emph{path} is a trail with no repeated nodes. 
 The \emph{distance} between two nodes is given by the number of edges of their shortest possible path. Considering a jet graph after $L$ message-passing operations, any two nodes with a distance less than or equal to $L$ would have information about each other encoded in the updated node features. A \emph{graph cycle} is a trail where the first and the last node corresponds to the same node, with all other nodes distinct. A \emph{cyclic graph} has at least one graph cycle. If a graph has no graph cycles, it is called an \emph{acyclic graph}. A connected acyclic graph is called a \emph{tree}. QCD splittings are naturally described by a tree~\cite{Andersson:1988gp,Dreyer:2018nbf}. 
 
A \emph{simple graph} is one where two distinct nodes can have at most one connection, and there are no self-loops. A graph is \emph{undirected} if we do not distinguish an edge based on the order of the nodes it connects; instead of the edge being defined by an ordered pair, we define it by an unordered pair. Constructing a directed graph will inherently have richer structural information of the underlying space on a point cloud.  If a graph is simple, it can be equivalently represented in terms of \emph{neighbourhood sets} $\mathcal{N}(i)$ in place of the edge set $\mathcal{E}$. For a directed graph, $\mathcal{N}(i)$ is defined for each node $i$, as the set containing all the nodes with incoming connections to $i$. We can allow for self-loops if we take the \emph{closed neighbourhood} where $i$ is also a part of the set $\mathcal{N}[i]\ni i$. We will see that allowing for self-loops is necessary for an IRC safe definition of message-passing operation on the graph. The $l$-hop neighbourhood of a node $i$ is the set of all nodes with distances from $i$, less than, or equal to $l$. In the rest of the work, we interchangeably use nodes or edges and their respective sets of representation in an underlying space.

\subsection{Message-passing neural networks}

Modern deep neural networks (DNNs) generally have a two-stage architecture: a specialised feature extraction section, followed by a generic dense architecture processing the extracted information further. Message-passing neural networks (MPNNs) are specialised feature extraction modules that act on graphs with node features and edge features. A message-passing operation, takes as input a graph with node features $\mathbf{h}^{(l)}_i$ and updates it to $\mathbf{h}^{(l+1)}$, with a two-step process:
\begin{enumerate}
	\item \textbf{Message passing:} We define a learnable function $\Phi^{(l)}$ with trainable parameters\footnote{This can be a multilayer perceptron, but it can be a collection of sub networks arranged in some particular way in general.}, which takes as input the node features $\mathbf{h}^{(l)}_i$ and $\mathbf{h}^{(l)}_j$, connected by the edge\footnote{It can also take the corresponding edge feature $\mathbf{e}^{(l)}_{ij}$ if present.} $(i,j)$ and returns the message ${}^i\mathbf{m}^{(l)}_{j}$,
	\begin{equation}
	\label{eq:msg_pass}
	{}^i\mathbf{m}^{(l)}_{j}=\Phi^{(l)}(\mathbf{h}^{(l)}_i,\mathbf{h}^{(l)}_j)\quad.
	\end{equation}
	The message is calculated for all edges present in the graph. We choose to use the notation $^i\mathbf{m}^{(l)}_j$ instead of a homogenous subscript, to make it evident that the message or the function $\Phi^{(l)}$ need not be symmetric with respect to the source node $j$, and the destination node $i$. 
	\item \textbf{Node-readout:} The nodes are updated with new features $\mathbf{h}^{(l+1)}_i$ by applying a permutation invariant function $\Box^{local}$ to all incoming messages \begin{equation}
	\label{eq:node_readout}
	\mathbf{h}^{(l)}_i\to\mathbf{h}^{(l+1)}_i=\Box^{local}_{j\in\mathcal{N}(i)}{}^i\mathbf{m}^{(l)}_{j}\quad.
	\end{equation} 
	Note that the updated features $\mathbf{h}^{(l+1)}_i$, contain the \emph{local neighbourhood} information of $i$.
\end{enumerate}
\begin{figure}[t]
	\centering 
	\includegraphics[scale=0.7]{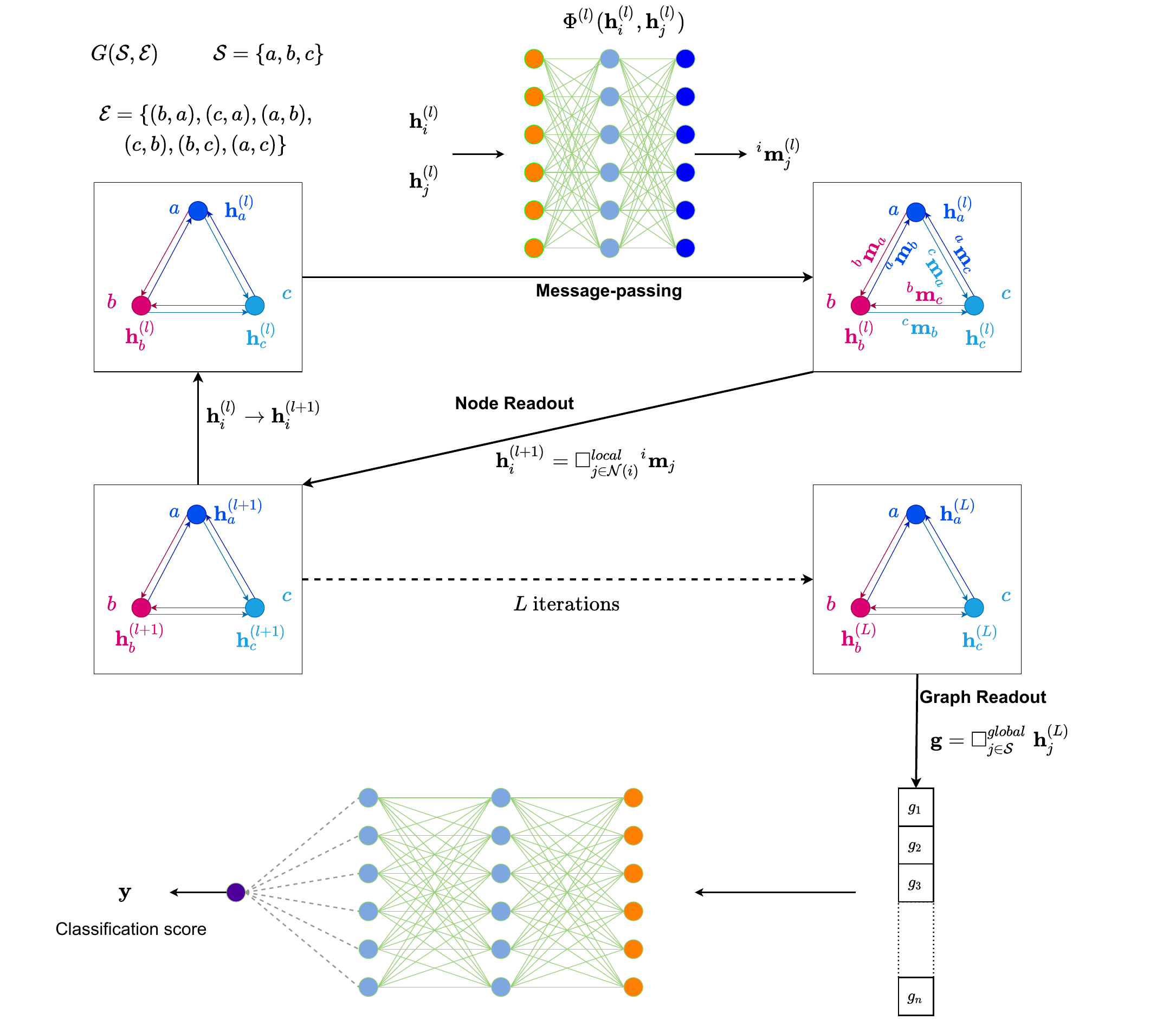}
	
	\caption{A diagramatic representation of a \emph{message-passing neural network}(MPNN) for \emph{graph-classification}. We are given a graph $G(\mathcal{S},\mathcal{E})$ with nodes $\mathcal{S}$ and edge set $\mathcal{E}$. Each node $i$ has a feature vector $\mathbf{h}^{(l)}_i$. The first step called the \emph{message-passing} involves evaluating the message ${}^i\mathbf{m}^{(l)}_j$ for each edge in $(j,i)$ via a DNN $\Phi^{(l)}$ shared for all edges. The different MPNN proposed in literature has structural differences in how $\Phi^{(l)}$ takes the inputs, which could include edge features as well. The second step, called the node readout, updates the features of each node to $\mathbf{h}^{(l+1)}_i$ with a permutation invariant function $\Box^{local}$ acting on all incoming messages. After $L$ iterations, a graph readout function $\Box^{global}$ on the final node features $\mathbf{h}^{(L)}_i$, gives fixed length $n$-dimensional graph representation $\mathbf{g}$. This is fed into a downstream neural network which outputs the graph classification score $\mathbf{y}$. }
	\label{fig:mpnn} 
\end{figure}  
The message-passing operation can be repeated any number of times. Each iteration leads to a gradual increase in the neighbourhood information held within the node features. On a \emph{static graph} where the neighbourhood sets $\mathcal{N}(i)$ or equivalently the edge set $\mathcal{E}$ remain unchanged, the node features contain information of the $L$-hop neighbourhood when applied $L$ times. Thus, the number of message-passing operations applied $L$, is a crucial hyperparameter in any GNN. It determines the scale at which the final node features $\mathbf{h}^{(L)}_i$, capture the \emph{local structures} within the graph. The number $L$ is restrained by the high computational cost of applying message passing operations, thereby reducing expressive power for the classification of large graphs. Even for jet graphs, we have a relatively large number of nodes, and hence, the information gets restricted to a local scale, intrinsically determined by $L$. For instance, in a two-prong $W$ tagging case, if $L$ is lesser than the length of the path between the two hard subjets, which would vary for each jet graph, the message-passing functions $\Phi^{(l)}$ would not see this feature for jets with several soft particles $n_{soft}>L$, between the two subjets. Nevertheless, a precisely determined graph construction algorithm would probabilistically give graphs with very low $<n_{soft}>\approx 0$. To avoid this limitation in the message-passing step \emph{dynamic} graphs are used to gather information from different scales, with the possibility of learning correlations from the entire graph after one dynamic iteration. However, for the same $L$, dynamic MPNNs will have a higher computational cost because of the added graph construction after each message-passing operation. Moreover, it may not always be desirable to mix information at the message-passing stage as the graph representation would still have the global features intact. Note that the number of nodes in a graph can vary. For graph classification, a permutation invariant graph readout function $\Box^{global}$ is applied to these node features, giving a fixed-length \emph{global representation} of the graph
\begin{equation}
\label{eq:graph_readout} 
\mathbf{g}=\Box^{global}_{i\in G} \;\mathbf{h}^{(L)}_i . 
\end{equation} In all instances, a graph readout serves similar purposes to the node readout, with the only difference being the scale of the operation. The graph representation is fed into a densely connected network, which outputs a classification score for the whole graph. The steps of an MPNN for graph classification, which we have discussed, are shown diagrammatically in Figure~\ref{fig:mpnn}. 

\section{IRC safe message-passing}
\label{sec:ewpm}
In this section, we examine the subtleties of building an IRC safe message-passing neural network. We can divide this into three steps: graph construction, message-passing and node readout, and graph readout.
In the following, we analyse the graph construction algorithm in Section~\ref{sec:irc_graph}, and the message-passing, node readout and graph readout together in Section~\ref{sec:ewpm_sub}. 

\subsection{Constructing the neighbourhood of a particle}
\label{sec:irc_graph}

\begin{figure}[t]
	\centering 
	\includegraphics[scale=0.8]{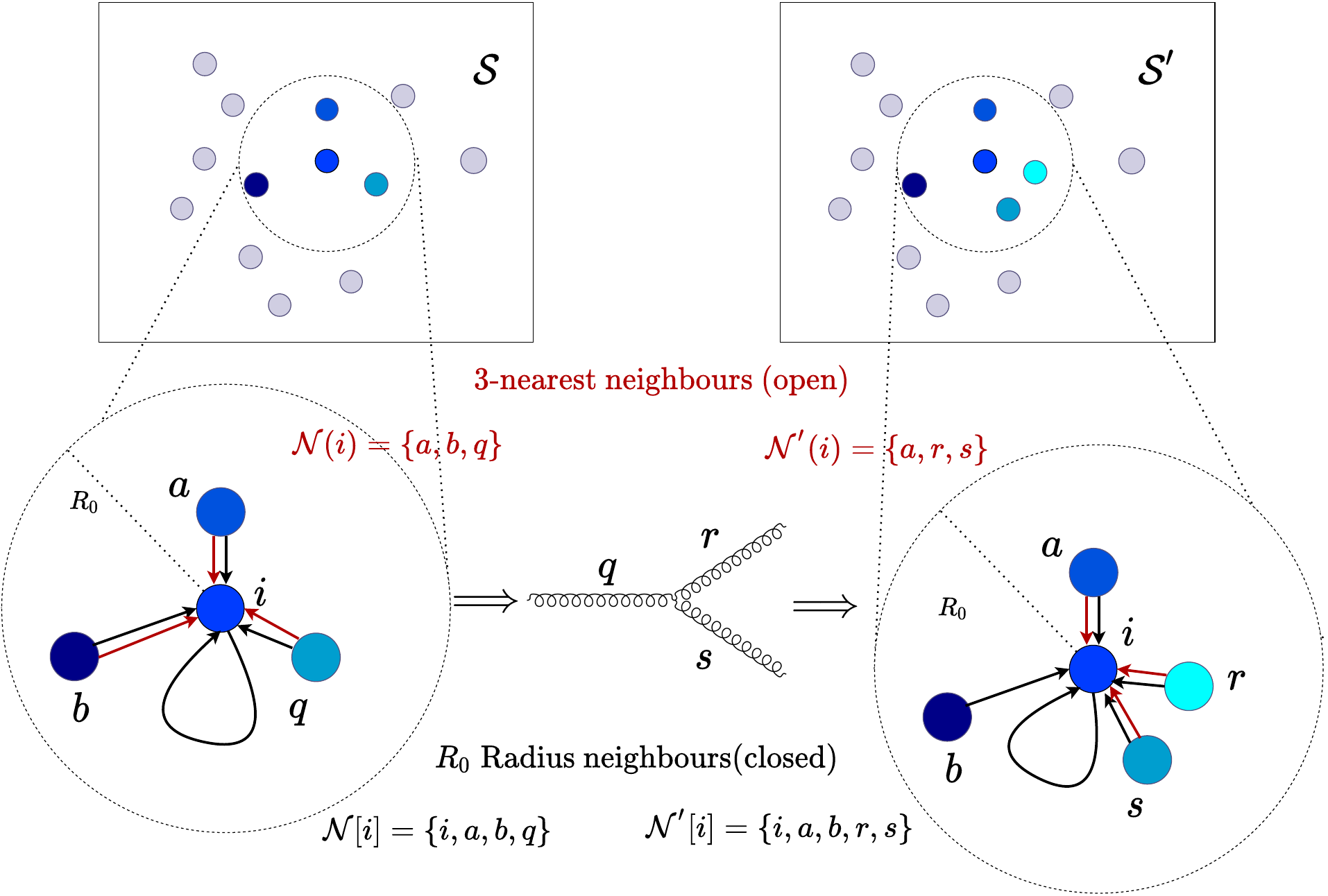}
	
	\caption{A $k$-nearest neighbour graph in the $(\eta,\phi)$-plane will have a different structure when any particle $q$ splits to $r$ and $s$. The set $\mathcal{S}$ denote the particles in the jet when there is no splitting, while $\mathcal{S}'$ denotes the particles with $q$ splitting. We show the directed edge connection to $i$ from its three nearest neighbours with red on either side. The neighbourhood set $\mathcal{N}(i)$ has $b$ in it, however when $q$ splits, $\mathcal{N}'(i)$ does not contain $b$. Therefore, the graph's structure prevents a smooth extrapolation between the two scenarios in the infra-red and collinear limit. This is not the case for a radius graph with radius $R_0$ in the $(\eta,\phi)$ plane, which is shown with black connections. We also include the self-loop of $i$, by using the closed neighbourhood sets $\mathcal{N}[i]$ and $\mathcal{N}'[i]$, since the node $i$ could also split into two particles. }
	\label{fig:knn_fail} 
\end{figure}

An infra-red and collinear safe observable has to be equal in the presence or absence of soft or collinear particles. Specifically, given a set $\mathcal{S}$ of $n$ massless particles with their four momenta $p_i=(z_i,\hat{p}_i)$, with $z_i=p^i_T/\sum_{j\in\mathcal{S}} p^j_T$ denoting the relative hardness of the particle, and $\hat{p}_i$ being the directional (angular) coordinates. If a particle $q$ undergoes a splitting $q\to r+s$, with $p_q=p_r+p_s$, an IRC safe observable $\mathcal{O}_n$ must satisfy
\begin{equation}
\label{eq:irc_def}
\begin{split} 
\mathcal{O}_{n+1}(p_a,...,p_b,p_r,p_s,p_c,...)&\to \mathcal{O}_{n}(p_a,...,p_b,p_q,p_c,...) \quad\text{as}\quad z_r\to 0\quad,\\
\mathcal{O}_{n+1}(p_a,...,p_b,p_r,p_s,p_c,...)&\to \mathcal{O}_{n}(p_a,...,p_b,p_q,p_c,...) \quad\quad\text{as}\quad \Delta_{rs}\to 0\quad,
\end{split} 
\end{equation} 
where $z_r$ is the relative hardness of $p_r$, and $\Delta_{rs}$ is the angle between $\vec{p}_r$ and $\vec{p}_s$. Consequently, the algorithm for constructing graphs should allow for the addition of soft or collinear particles without changing the whole structure of the graph. The graph constructed by a vertex deletion of a soft or collinear particle should be equal to the one formed in its absence, with proper substitution of the four-momenta in the case of collinear particles. For instance, a k-nearest neighbour (k-NN) graph would not allow for an IRC safe message-passing since adding a particle in the vicinity of a node $i$ could change the neighbourhood set $\mathcal{N}(i)$ to $\mathcal{N}'(i)$ with a fixed cardinality. The fixed cardinality would induce a \emph{domino effect} in the neighbourhood sets of the subsequent neighbours and change the graph's structure to a large degree. As a concrete example, for a k-NN graph in the $(\eta,\phi)$ plane, the addition of a particle closer to the node could, in principle, omit the hardest particle out of the neighbourhood in $\mathcal{N}'(i)$. This is diagrammatically shown in Figure~\ref{fig:knn_fail}, where a particle $q$ splitting to two particles $r$ and $s$ excludes another particle $b$ from the neighbourhod of particle $i$. Thus, in the node readout for particle $i$, a message-passing algorithm based on a k-NN graph cannot smoothly extrapolate between the two scenarios, when taking the IRC limits of the daughter particles $r$ and $s$. This warrants a careful examination of the graph construction algorithm. 

Since our final aim is to have a message-passing neural network whose output is IRC safe, the correctness of the graph construction algorithm is intimately connected with the subsequent operations the network will perform on the graph's nodes. From the perspective of QCD, the node readout and the graph readout functions are on the same footing, with the only difference being the scale. We look into the jet substructure with the help of the nodes and the edge connections, which gives us a representation of the whole jet\footnote{This could be extrapolated to the event shape, where an IRC safe graph neural network would look into the subsequent scales present in the event and construct an event level representation which would have the desirable property of being less affected by soft and collinear radiations.}. In graph theory, self-loops are often ignored, and most of the efforts concentrate on analysing simple graphs. However, from the perspective of QCD, the destination node itself can also emit soft or collinear particles. Therefore, an IRC safe aggregation must act on the closed neighbourhood $\mathcal{N}[i]$, which includes the destination node $i$. 

Let us take a set $\mathcal{S}$ of the four momenta of $n$ massless particles. Out of these, any particle $q$ could undergo a splitting to $r$ and $s$, which enlarges the set $\mathcal{S}$ to $\mathcal{S}'$ with $\mathcal{S}'=\mathcal{S}\setminus\{q\}\cup \{r,s\}$. The three four momenta can be written in general as 
\begin{equation} 
p_q=(z_q,\hat{p}_q)\quad,\quad p_r=(z_r,\hat{p}_r)\; \; z_r=\lambda z_q\;\quad,\quad p_s=(z_s,\hat{p}_s),\;\;z_s=(1-\lambda)\;z_s \quad,
\end{equation} with $\lambda\in[0,1]$, and $p_q=p_r+p_s$. Following are the limits that are of interest:
\begin{itemize} 
	\item\textbf{IR limit:} $\lambda\to 0$($\lambda\to 1$), for $r$(or $s$) in the soft limit, 
	\item\textbf{C limit:} $\hat{p}_r\to\hat{p}_s\to\hat{p}_q$ or equivalently $\Delta_{rs}\to0$, for any $\lambda$.
	\end{itemize} 
	For the IR limit, the two cases are for either of the daughter particles becoming soft, and it suffices to take one of them, say $\lambda\to0\implies z_r\to0$ in the following presentation.
	A graph construction method on $\mathcal{S}$ would allocate to each particle $i$ a neighbourhood set $\mathcal{N}[i]\subseteq \mathcal{S}$. We would have to apply the same method to $\mathcal{S}'$, which would give neighbourhood sets $\mathcal{N}'[i]\subseteq \mathcal{S}'$. To devise an IRC safe message passing operation, a simple procedure is to assume that the neighbourhood sets, $\mathcal{N}[i]$, behave the same way as the total set $\mathcal{S}$. By keeping the behaviour of the sets the same, the graph structure essentially works as a control over the scale of the message-passing operation. In the IR limit, the emitter $q$ and the daughter $r$ need not fall in the same neighbourhood since the other daughter $s$ will have the same four momenta of $q$ in the limit $z_r\to0$. However, in the C limit, if the emitter $q$ is in the neighbourhood of $\mathcal{N}[i]$, we need both the daughters to be in $\mathcal{N}'[i]$. Mathematically, we can write this condition as:
	
	\begin{itemize}  
	\item \textbf{IR limit:} If $r\notin \mathcal{N}'[i]\implies \mathcal{N}'[i]=\mathcal{N}[i]$ 
\begin{equation} 
\label{eq:neighbour_set_ir}
\text{else}\;\mathcal{N}'[i]\setminus\{r\}=\mathcal{N}[i]\quad, \text{when }\; z_r=0\;;
\end{equation} 
\item\textbf{C limit:} If $\{r,s\}\cap\mathcal{N}'[i]=\emptyset \implies \mathcal{N}'[i]=\mathcal{N}[i]$  
\begin{equation} 
\label{eq:neighbour_set_c} 
\text{else}\;\mathcal{N}[i]= \mathcal{N}'[i]\setminus\{r,s\}\cup\{q\}\quad,\text{when }\;\Delta_{rs}=0\;.
\end{equation} 
\end{itemize} 

The aim now is to devise a graph construction algorithm that will give us neighbourhood sets satisfying these conditions. Constructing graphs from sets sampled from a point cloud uses functions defined on the features $\vec{\beta}_i$. The algorithm can be surmised by comparing two functions, which, in general, depend on features $\vec{\beta}_i$ (which need not be the same as the node features $\mathbf{h}_i$) of elements $i$ belonging to subsets of the whole sample set $\mathcal{S}$, which itself can change as the edge set $\mathcal{E}$ grows. Calling these two functions as the \emph{decision} function $\mathbf{D}$, and the \emph{threshold} function $\mathbf{T}$, we say that a particle $j$ will be placed into the neighbourhood of $i$, if $\mathbf{D}$ is less than or equal to $\mathbf{T}$, 
\begin{equation}
\label{eq:graph_algo}
\mathbf{D}(\vec{\beta}_i,\vec{\beta}_j|\vec{\beta}_k,\vec{\beta}_l,...)\leq\mathbf{T}(\vec{\beta}_i,\vec{\beta}_j|\vec{\beta}_k,\vec{\beta}_l,...)\implies j \in \mathcal{N}[i]\quad. 
\end{equation} The features $\vec{\beta}_i$ can generally contain any quantity of $i$ like charge, four-momenta, or the identity of the sub-detector component of $i$. Graphs are versatile data structures that can encode the detector components together into a compact, unified representation. However, as our current aim is to incorporate IRC safety, it restricts us to calorimeter or particle flow constituents with no charge information and the four vectors of the particles. In the following, we systematically reduce the possible four-vectors which could come into the arguments of the decision and the threshold functions. 

As was previously discussed, $\mathbf{D}$ or $\mathbf{T}$ cannot be dependent on the cardinality of the neighbourhood set $\mathcal{N}[i]$. Consider the functions depending on another particle $p_q$ to decide whether $p_j$ should be in $\mathcal{N}[i]$. A splitting on $p_q$ can create situations where $p_j$ can be in $\mathcal{N}[i]$ and not in $\mathcal{N}'[i]$ or vice versa. Thus, the functions can not depend on any other four vectors than the two particles in question. Looking at Eq.~\ref{eq:neighbour_set_c}, we see that the emitter and the daughter particles of a collinear splitting need to be in both in the neighbourhood $\mathcal{N}[i]$  and $\mathcal{N}'[i]$, respectively, or not at all. We have the following condition on the decision and threshold functions,
\begin{equation}
\label{eq:DT_c_split_dest}
\begin{split}
\mathbf{D}(p_i,p_r+p_s)\leq \mathbf{T}(p_i,p_r+p_s)\Leftrightarrow \mathbf{D}(p_i,p_r)\leq \mathbf{T}(p_i,p_r)\;\text{and}&\;\mathbf{D}(p_i,p_s)\leq \mathbf{T}(p_i,p_s)\quad,\\
\mathbf{D}(p_r+p_s,p_i)\leq \mathbf{T}(p_r+p_s,p_i)\Leftrightarrow \mathbf{D}(p_r,p_i)\leq \mathbf{T}(p_r,p_i)\;\text{and}&\;\mathbf{D}(p_s,p_i)\leq \mathbf{T}(p_s,p_i)\quad,
\end{split}  
\end{equation}
in the exact collinear limit of $\Delta_{rs}= 0$. The second line arises when considering the emitter or daughters as the destination node, with $p_i$ denoting any particle in their respective sets. A simple way to satisfy these inequalities is by using the condition of collinearity and making the functions dependent only on the directional coordinates,  
\begin{equation}
\mathbf{D}=\mathbf{D}(\hat{p}_i,\hat{p}_j)\quad,\quad \mathbf{T}=\mathbf{T}(\hat{p}_i,\hat{p}_j)\quad.
\end{equation} 
The functions can also have additional dependence on any IRC safe quantity defined on the set $\mathcal{S}$. 

 For our network analysis, we explore the simplest possible graphs to gauge the power of this method by constructing graphs with constant radius $R_0$, 
 \begin{equation}
 \label{eq:use_DT}
 \mathbf{D}=\Delta R_{ij}\quad,\quad \mathbf{T}=R_0\quad, 
 \end{equation} in the $(\eta,\phi)$-plane. Complicated dependencies on the directional variables and on IRC safe quantities like the jet's $p_T$ can be explored in future work. The black connections to particle $i$ in Figure~\ref{fig:knn_fail} show a case where a split in particle $q$ preserves the other particles in the neighbourhood sets, except for the emitter and the daughters.

\subsection{Energy-weighted Message-Passing}
\label{sec:ewpm_sub}
Now that we have the graph construction algorithm, we look into building an IRC safe message-passing function. The message-function at the first layer $\Phi^{(0)}$ would take two four-vectors $p_i$ and $p_j$ for a \emph{directed edge} from $j$ to $i$, to give the message ${}^i\mathbf{m}^{(0)}_j$. The node features are then updated to $\mathbf{h}^{(1)}_i$, by applying a permutation invariant function on the messages ${}^i\mathbf{m}^{(0)}_j$ for all possible $j\in\mathcal{N}[i]$. 
Commonly used permutation invariant functions can be classified in the sense of QCD into \emph{exclusive} or \emph{inclusive} functions: the function output depends on a specific subset of the neighbourhood, or it depends equally on all the neighbourhood particles. Max/min falls within the first class, while mean/sum falls under the second class. As one can presume, it is inherently problematic to build IRC safety into exclusive functions. Building IRC safety into a mean readout is not straightforward since it depends explicitly on the number of particles in $\mathcal{N}[i]$. 
In the following, we examine the conditions which give IRC safety of the updated node features on the message-passing function for the exclusive and summed node readout operations.  

\vspace{0.2cm}
\noindent
\textbf{Max/Min readout:} Since, the only difference between max and min readout is the comparison, we look at max readout. The same for min readout follows by replacing the greater-than with the less-than symbol in the message comparisons. We have the messages $\Phi^{(0)}(p_i,p_j)={}^i\mathbf{m}^{(0)}_j$ with the max update as, $$\mathbf{h}^{(1)}_i=\max_{j\in \mathcal{N}[i]} \Phi^{(0)}(p_i,p_j)\quad.$$
For a splitting $q$ to $r$ and $s$ with $p_q= p_r+ p_s$ and assuming that the neighbourhood sets follow Eq.~\ref{eq:neighbour_set_ir} and \ref{eq:neighbour_set_c}. 
In the soft limit when $z_r\to0$, we have $$z_j> z_r\implies \Phi^{(0)}(p_i,p_j)> \Phi^{(0)}(p_i,p_r)\quad\text{for IR safety.}$$  For the collinear limit $\Delta_{rs}\to0$, we have $$\Phi^{(0)}(p_i,p_j)\geq\Phi^{(0)}(p_i,p_r) \quad\text{and}\quad \Phi^{(0)}(p_i,p_j)\geq\Phi^{(0)}(p_i,p_s)\;\forall\; j \;\in\; \mathcal{N}[i]\quad\text{for C safety.}$$  Implementing C safety in a max/min node readout is \emph{not possible} since the angle $\Delta_{rs}$ needs to control the ordering of the messages ${}^i\mathbf{m}^{(0)}_r$ and ${}^i\mathbf{m}^{(0)}_s$ with all other messages ${}^i\mathbf{m}^{(0)}_j$. The max function chooses the maximum value out of all ${}^i\mathbf{m}^{(0)}_j$, with the ordering essentially determined by the second argument in $\Phi^{(0)}$. Consider an exactly collinear splitting of the particle contributing to the highest message vector in $\mathcal{N}[i]$, say $p_M\to \lambda\;p_M+(1-\lambda)\;p_M$, with $\lambda\in[0,1]$. The max value in both the scenarios can be equal only at the endpoints $\lambda\in\{0,1\}$, which is essentially the soft \emph{and} collinear limit. The same is true for min readout when considering the particle determining the minimum value of $\Phi^{(0)}$ in the neighbourhood.

\vspace{0.2cm}
\noindent
\textbf{Sum readout:} The updated node features are given by,	
\begin{equation} 
\label{eq:sum_pool}
\mathbf{h}^{(1)}_i=\sum_{j\in\mathcal{N}[i]}\Phi^{(0)}(p_i,p_j)\quad .
\end{equation}
For a splitting $q\in\mathcal{N}[i]$ to $r,s\in\mathcal{N}'[i]$ changing the neighbourhood set $\mathcal{N}[i]$ to $\mathcal{N}'[i]$. The requirements on the message function $\Phi^{(0)}$ are \addtocounter{equation}{-1}
\begin{subequations}
	\label{eq:irc_safe_msg}
\begin{align}\label{eq:ir_safe_msg}
&\text{\textbf{IR safety: }}&&\Phi^{(0)}(p_i,p_r)\to 0 \quad\text{as}\quad z_r\to0\quad\\
\label{eq:c_safe_msg}
&\text{\textbf{C safety: }}&&\Phi^{(0)}(p_i,p_r+p_s)= \Phi^{(0)}(p_i,p_r)+\Phi^{(0)}(p_i,p_s)\quad \text{as} \quad \Delta_{rs}\to 0\quad.
\end{align} 
\end{subequations}Satisfying these conditions gives IRC safe updated node features 
\begin{equation} 
\label{eq:irc_safe_node}\mathbf{h}^{(1)}_i=\mathbf{h}'^{(1)}_i=\sum_{j\in\mathcal{N}'[i]}\Phi^{(0)}(p_i,p_j)\quad.
\end{equation} 
We have written these conditions for the second argument only, even though the splitting can occur in the destination node, since it has a special status in the message passing operation. Applying similar conditions for the first argument is highly restrictive with no practical gain. Nodes corresponding to the daughters are present in the graph even in the IRC limit, which is precisely the objective of the present study -- to get fixed-length representations of two graphs, with one having an additional node, which is the same when the additional node or particle is soft or collinear. Including the destination node in the neighbourhood set makes it possible for the emitter and the two daughters to have the same updated node features in the exact collinear limit, with the two collinear copies propagating forward simultaneously. In either of the limits (soft or collinear), these copies are then taken care of separately by an IRC safe graph readout. These are explained in more detail in the following paragraphs.  

We now present an implementation of message-passing operation which satisfies the IRC safe conditions for a summed node readout. The message function $\Phi^{(0)}$ has a dependence on two four-vectors, which allows an MPNN to extract richer features than the ones employed in EFNs~\cite{Komiske:2018cqr} with a single particle map. However, the per-particle map can be functionally regarded as a special message function constant for the second argument. The point cloud could then be regarded as a graph of $N$ nodes, with $N$ disconnected components with only self-loops entering the edge set. Therefore, generalising the per-particle map, we define the message function as 
\begin{equation}
\label{eq:irc_msg} 
{}^i\mathbf{m}^{(0)}_{j}=\Phi^{(0)}(p_i,p_j)=\omega^{(\mathcal{N}[i])}_j\;\;\hat{\Phi}^{(0)}(\hat{p}_i,\hat{p}_j)\quad,
\end{equation} where $\hat{\Phi}$ takes only the directional information of the four vectors and we define scalar weights $\omega^{(\mathcal{K})}_j$, dependent on the scope $\mathcal{K}$ of the readout operation,
\begin{equation}
\label{eq:weight_scope}
\omega^{(\mathcal{K})}_j=\frac{p^j_T}{\sum_{k\in\mathcal{K}}\; p^k_T}\quad.
\end{equation}
Clearly, for the full set $\mathcal{S}$, $\omega^{(\mathcal{S})}_j=z_j$, and $z_j\to0\implies\omega^{(\mathcal{K})}_j\to0$ regardless of $\mathcal{K}$, thereby satisfying\footnote{Note, when a soft particles has no other neighbour except itself, the node readout might change to a finite value. However, the graph readout, and therefore the network output, will remain unchanged, as $\mathcal{K}=\mathcal{S}$.} Eq.~\ref{eq:ir_safe_msg}. Moreover, as long as the neighbourhood sets $\mathcal{N}[i]$ and $\mathcal{N}'[i]$ satisfy Eq.~\ref{eq:neighbour_set_c} which is true even when $i$ undergoes a splitting we have,
$$\omega^{(\mathcal{N}[i])}_q=\omega^{(\mathcal{N}'[i])}_r+\omega^{(\mathcal{N}'[i])}_s\quad,$$ where $q$ is the emitter and $r$ and $s$ are the daughter particles. Since $\hat{p}_q=\hat{p}_r=\hat{p}_s$ in the collinear limit, we have \begin{equation} \label{eq:coll_msg} 
\begin{split} 
\hat{\Phi}^{(0)}(\hat{p}_{i},\hat{p}_{q})\;=&\;  \hat{\Phi}^{(0)}(\hat{p}_{i},\hat{p}_{r})\;=\; \hat{\Phi}^{(0)}(\hat{p}_{i},\hat{p}_{s})\quad,\\
\hat{\Phi}^{(0)}(\hat{p}_{q},\hat{p}_{i})\;=&\; \hat{\Phi}^{(0)}(\hat{p}_{r},\hat{p}_{i})\;=\; \hat{\Phi}^{(0)}(\hat{p}_{s},\hat{p}_{i},)\quad.
\end{split} 
\end{equation} Hence, the updated node features 
\begin{equation}
\label{eq:ewmp_inp}
\mathbf{h}^{(1)}_i=\sum_{j\in\mathcal{N}[i]}\;\omega^{(\mathcal{N}[i])}_j\;\;\hat{\Phi}^{(0)}(\hat{p}_i,\hat{p}_j)\quad,
\end{equation} satisfies the IRC safety condition Eq.~\ref{eq:irc_safe_node}. Note that the expression does not limit the form of the function $\hat{\Phi}^{(0)}$ other than differentiability which is required for back-propagation. Thus, we can modify any existing message-passing algorithm into the IRC safe version by implementing the appropriate message weights and restricting the input to the directional coordinates. We therefore implement the IRC safe version of edge-convolutions as a proof-of-principle analysis.  

Looking at the structure of the updated node features after the first message-passing operation, we can see that it is a function of all the four-momenta of its neighbourhood particles. If $n$ is the number of nodes in the set $\mathcal{N}[i]$, we have the updated IRC safe node feature as $\mathbf{h}^{(1)}_i(p_1,p_2,...p_n)$. We want to investigate the IRC safety of another message passing on the updated quantities $\mathbf{h}^{(1)}_i$. If true, the architecture could accommodate multiple iterations of message-passing operations, thereby increasing the model's expressive power. For simplicity, one can consider static graphs with the same neighbourhood sets. A weighted message-passing with of the same form as Eq.~\ref{eq:ewmp_inp} $$\mathbf{h}^{(2)}_i=\sum_{j\in\mathcal{N}[i]}\;\omega^{(\mathcal{N}[i])}_j\;\hat{\Phi}^{(1)}(\mathbf{h}^{(1)}_i,\mathbf{h}^{(1)}_j)\quad,$$ with the same weights $\omega^{(\mathcal{N}[i])}_j$, but with the updated node features $\mathbf{h}^{(1)}_i$ satisfies IR safety. For it to be C safe, the features $\mathbf{h}^{(1)}_i$ should behave just like the directional coordinates $\hat{p}_i$. Note that the neighbourhood sets for the two collinear daughters are the same. The emitter also has the same neighbourhood after replacing the daughters with their summed four-vector (cf. Eq.~\ref{eq:neighbour_set_c}). Their aggregated node vectors become equal to that of the emitter in $\mathcal{S}$ via the cancellation of the $\lambda$ factors in the weights. Thus, the updated node vectors after the first message-passing of the daughters and the emitter are exactly equal in the collinear limit $\mathbf{h}^{(1)}_q=\mathbf{h}^{(1)}_r=\mathbf{h}^{(1)}_s$. Hence, they have essentially the same characteristics as the directional coordinates. This ensures that $\hat{\Phi}^{(1)}(\mathbf{h}^{(1)}_i,\mathbf{h}^{(1)}_j)$ follow analogous equations to Eq.~\ref{eq:coll_msg}, thereby making the weighted message $\omega^{(\mathcal{N}[i])}_j\;\hat{\Phi}^{(1)}(\mathbf{h}^{(1)}_i,\mathbf{h}^{(1)}_j)$ follow similar equations to Eq.~\ref{eq:irc_safe_msg}. Moreover, the new features $\mathbf{h}^{(2)}_i$, would have this same property. Hence, \emph{repeating the energy weighted message passing operation any number of times satisfies IRC safety at the level of each updated node feature.} Denoting the node features for the $l^{th}$ iteration as $\mathbf{h}^{(l)}_i$ with $\mathbf{h}^{(0)}_i=\hat{p}_i$, we have the iterative application of the energy-weighted message passing as 
\begin{equation}
\label{eq:ewmp}
\mathbf{h}^{(l+1)}_i=\sum_{i\in\mathcal{N}[i]}\;\omega^{(\mathcal{N}[i])}_{j}\;\;\hat{\Phi}^{(l)}(\mathbf{h}^{(l)}_i,\mathbf{h}^{(l)}_j)\quad.
\end{equation}

\begin{figure}[t]
 	
 	\includegraphics[scale=0.8]{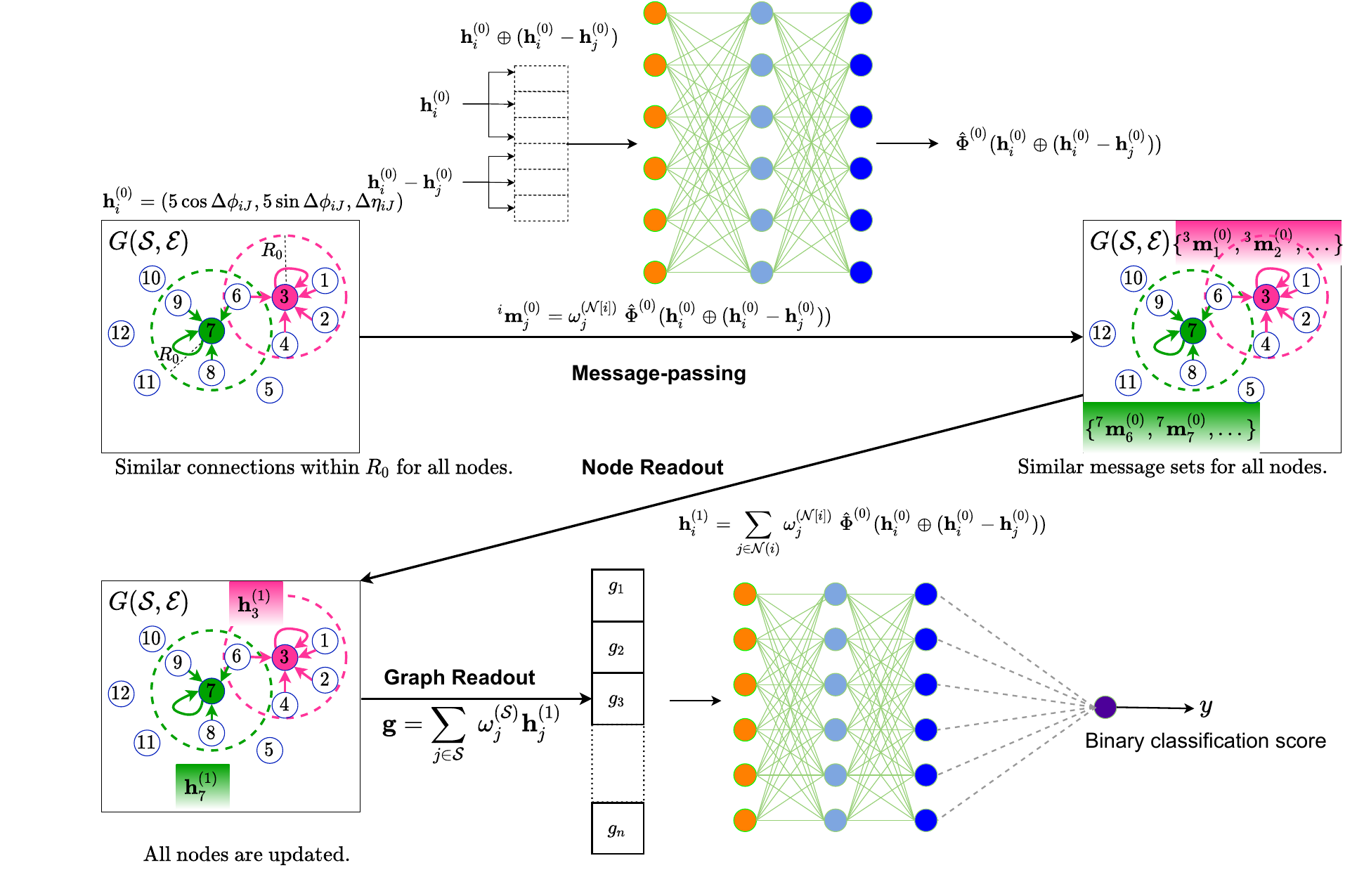}
 	\caption{The specific architecture used for the three jet tagging scenarios of an \emph{Energy-weighted Message-Passing network}(EMPN), with a single energy-weighted message passing operation. It takes graphs of constant radius $R_0$ in the $(\eta,\phi)$-plane. The message-passing network $\Phi^{(0)}$, takes the directional inputs of the four-vectors in the form of $\mathbf{h}^{(0)}_i$, and calculates a weighted message ${}^i\mathbf{m}^{(0)}_j$ with $\omega^{(\mathcal{N}[i])}_j$ as the weights. It then undergoes a summed node readout operation to update their features to $\mathbf{h}^{(1)}_i$. The graph representation $\mathbf{g}$ obtained after a summed graph readout operation on the node features $\mathbf{h}^{(1)}_j$ weighted with $\omega^{(\mathcal{S})}_j$, is fed into a DNN which outputs a binary classification score. }
 	\label{fig:empn} 
 \end{figure}

As seen above, there will be copies of emitted particles even in the IRC limit, propagating forward in the graph formed after a soft or collinear splitting. Thus, any generically defined graph readout operation acting on the node features of the full graph will not be IRC safe. The graph readout should guarantee the equality of the obtained representation of the two graphs in the IRC limit, with one having an additional node. The node features at the final message-passing layer, say $\mathbf{h}^{(L)}_i$, will behave the same way as the directional variables, regardless of $L$, the number of message-passing iterations. Thus, a graph readout of the form 
\begin{equation}
\label{eq:irc_graph_readout}
\mathbf{g}=\sum_{i\in G} \omega^{(\mathcal{S})}_i\;\mathbf{h}^{(L)}_i\quad,
\end{equation} with $z_i=\omega^{(\mathcal{S})}_i$ , is IRC safe. This is an analogue of the sum over the per-particle representation employed in EFNs. The graph convolution operation now replaces the per-particle maps. The scale of the representation which undergoes the sum, which contains local structural information, is determined by the number of message-passing operations and the graph construction algorithm. A schematic representation of such a network for $L=1$ is shown in Figure~\ref{fig:empn}.
 
\section{Details of network implementation}
\label{sec:net_imp}
In this section, we present the numerical results of the IRC safe message passing neural network. The details of the datasets are given first, followed by the network hyperparameters and training aspects. 
\subsection{Analysis setup}
\begin{table}[t]
	\begin{center}
	\begin{tabular}{|c|c|c|c|c|c|}
		\hline
		Sl.No&Jet Class&Parton-level&MPI&Detector Simulation&Jet-radius\\
    \hline
		1.& Gluon&{\tt Pythia8}&Yes&No&0.4\\
    2.& Quark&{\tt Pythia8}&Yes&No&0.4\\	
    3.& QCD&{\tt Pythia8}&No&Yes&0.8\\
    4.& Top&{\tt Pythia8}&No&Yes&0.8\\
    5.& $W$&{\tt MadGraph5\_aMC@NLO}&No&Yes&0.8\\     	
		\hline
		\end{tabular} 
	\end{center} 
	\caption{A summary of the different classes of data used in the three classification scenarios. The $W$ data was generated for this study, while for the first four classes, we use publicly availaible datasets~\cite{komiske_patrick_2019_3164691,kasieczka_gregor_2019_2603256}. All datasets were showered and hadronised with {\tt Pythia8}, while the detector simulation was done with {\tt Delphes3}, with the default ATLAS card.}
	\label{tab:dataset} 
	\end{table} 
For assessing the power of \emph{Energy-weighted Message Passing Networks} (EMPN), we consider three scenarios: quark/gluon discrimination as a benchmark for IRC safe, supervised identification of normal radius quark jets from gluon jets, boosted $W$  vs QCD jet tagging as an example of two-prong tagging, and boosted top vs QCD jet tagging as an example of three-prong tagging. We use publicly available datasets for the quark vs gluon tagging~\cite{komiske_patrick_2019_3164691,Komiske:2018cqr}, and the top tagging scenarios~\cite{kasieczka_gregor_2019_2603256,Butter:2017cot}. These datasets were generated at 14 TeV center-of-mass energy proton-proton collisions in {\tt Pythia8}~\cite{Sjostrand:2014zea}. The parton level events were generated in {\tt Pythia 8.226} using the processes {\tt WeakBosonAndParton:qqbar2gmZg}
and {\tt WeakBosonAndParton:qg2gmZq} for the gluon and quark samples respectively. These events were showered with the default tunings of the shower parameters with multi-parton interactions (MPI) and hadronisation. All final state particles except neutrinos were clustered with {\tt FastJet 3.3.0}~\cite{Cacciari:2011ma} into anti-$k_T$~\cite{Cacciari:2008gp} jets of radius $R=0.4$, with no detector simulation. Jets are required to have $p_T\in[500,550]$ GeV and rapidity $|y|<2$. Parton level events for QCD jets and top jets in the top tagging dataset were generated with {\tt Pythia 8.2.15}. These were showered without MPI effects and passed through {\tt Delphes3}~\cite{deFavereau:2013fsa}, with the default ATLAS detector card. The particle-flow objects are clustered into anti-$k_T$ jets with $R=0.8$. The jets are required to have $p_T\in[550,650]$ GeV, with pseudorapidity $|\eta|<2$. For the top-jets, the parton-level top quark and its decay products were required to fall within $\Delta R=0.8$ of the reconstructed jet axis. QCD jets from this dataset was used for the $W$-tagging scenario. For the $W$ jets, we generated the parton level process {\tt p p $>$ w$^\pm$ z } in {\tt MadGraph5\_aMC@NLO}(v2.6.5)~\cite{Alwall:2014hca}, at 14 TeV proton-proton collisions, forcing the $W$ boson to decay hadronically, and the $Z$ boson to decay to neutrinos. Parton level cuts on the missing-transverse energy with $\slashed{E}_T>500$ GeV, and the pseudorapidity of the $W$ bosons, $|\eta_{w}|< 3$, were applied during the generation. Further downstream simulation of these partonic events was done by implementing the same configuration details of the top-dataset, including the jet-reconstruction and baseline selection criteria. We also matched the parton level $W$ and its decay products to be within $\Delta R=0.8$ of the reconstructed $W$ jet axis. Up to two-hundred hardest constituents within the jet were used to construct the graphs for the three large-radius jet tagging datasets. For all three scenarios, we have 1.2 million training, 400k validation, and 400k test jets.  
\subsection{Constructing the jet graphs}  
The jet graphs of each jet are constructed by taking their constituents. We calculate the interparticle distance $\Delta R_{ij}=\sqrt{\Delta\eta^2+\Delta\phi^2}$, in the $(\eta,\phi)$ plane. For each node $i$, we define the neighbourhood set $\mathcal{N}[i]$ as the set of all the particles $i$ with $\Delta R_{ij}\leq R_0$. After the neighbourhood sets, or equivalently the edge set $\mathcal{E}$ of the graph are obtained, we shift the coordinates of each constituents $(\eta_i,\phi_i)$ to $(\Delta\eta_{iJ},\Delta\phi_{iJ})$, their distance between the jet axis $(\eta_J,\phi_J)$. The node features that the network takes has the $\phi$ coordinates mapped to two-dimensional coordinates $(a\cos\phi,a\sin\phi)$. Keeping in mind the total allowed range of $\eta\in[-5,5]$, we choose $a=5$. Thus, for each jet constituent, we have the node features of the input graph as
$$\mathbf{h}^{(0)}_i=(5\cos\Delta\phi_{iJ},5\sin\Delta\phi_{iJ},\Delta\eta_{iJ})\quad.$$ This choice of representation makes the edge-convolution (which we will be using) look at the $\phi$ information through an \emph{embedding} in a two-dimensional Euclidean space. This is essential since multilayer perceptrons (MLPs), the building blocks of neural networks, are essentially sequential \emph{affine} maps interspersed by non-linear activation functions, and the periodicity of $\phi$ may not be evident to it directly even if the graph has the periodicity. The range of $\phi$ for jets considered here are not wide enough for the periodicity to become a major bottleneck. However, it is crucial when considering the inclusive event information. We also calculate the weights $\omega^{(\mathcal{K})}_{j}$ defined in Eq.~\ref{eq:weight_scope}, for all neighbourhood sets $\mathcal{N}[i]$ and the full set $\mathcal{S}$. The jets are not preprocessed with steps like rotation and reflection before extracting the node features. Doing so should improve the network performance as these symmetries are not built into the architecture. Incorporating these symmetries into the architecture could also improve the performance in the absence of preprocessing.  
\subsection{Network hyperparameters and training}

As this is a proof-of-principle study, we examine the simplest of architectures to showcase the ability of EMPNs at the different classification scenarios. We implement an {\tt EnergyWeighted} message-passsing module in {\tt PyTorch-Geometric-1.7.2}~\cite{Fey/Lenssen/2019}, for the analysis of the EMPN network. The message-passing function corresponds to an energy-weighted edge convolution~\cite{wang2019dynamic},
\begin{equation}
\label{eq:edge_conv}
{}^i\mathbf{m}^{(0)}_{j}=\omega^{(\mathcal{N}[i])}_j\;\hat{\Phi}^{(0)}\left(\mathbf{h}^{(0)}_i\oplus(\mathbf{h}^{(0)}_j-\mathbf{h}^{(0)}_i)\right)\quad.
\end{equation}The learnable function $\hat{\Phi}^{(0)}$ is an MLP having two hidden layers. The input layer takes the six-dimensional concatenated vector $\mathbf{h}^{(0)}_i\oplus(\mathbf{h}^{(0)}_j-\mathbf{h}^{(0)}_i)$, and maps it to a 128-dimensional representation. Both hidden layers are also fixed to have 128 nodes each with {\tt ReLU} activations, while the output layer has {\tt Linear} activation. The graph representation obtained after applying the IRC safe-readout (cf. Eq.~\ref{eq:irc_graph_readout}) is fed into a downstream MLP, which outputs the binary classification score. This MLP has three hidden layers, with all of them having sixty-four nodes and {\tt ReLU} activations. The structure of the EMPN network is summarised in Figure~\ref{fig:empn}. We use Adam~\cite{kingma2014adam} optimiser with an initial learning rate of 0.001, which reduces with a decay-on-plateau condition by a factor of 0.5, with the patience of two epochs without any cooldown. We scan over a set of $R_0$ values for each classification scenario. For the $W$ and top tagging with large-radius jets ($R=0.8$), we choose $R_0\in\{0.1,0.2,0.3,0.4,0.5\}$, while for the quark-gluon classification with normal-radius jets ($R=0.4$), we choose $R_0\in\{0.1,0.2,0.3,0.4\}$. For all three scenarios and each $R_0$, we train the same network from random initialisation five times. All networks were trained for seventy epochs. The epoch with minimum validation loss is used for evaluating the model with their respective test datasets for each instance of the training. Note that we do not perform any hyperparameter optimisation, and doing so should further improve the performance.  

\section{Results}
\label{sec:results}
\subsection{Tagging performance}
\begin{figure*}[t]
	\centering	
	
	\includegraphics[scale=0.35]{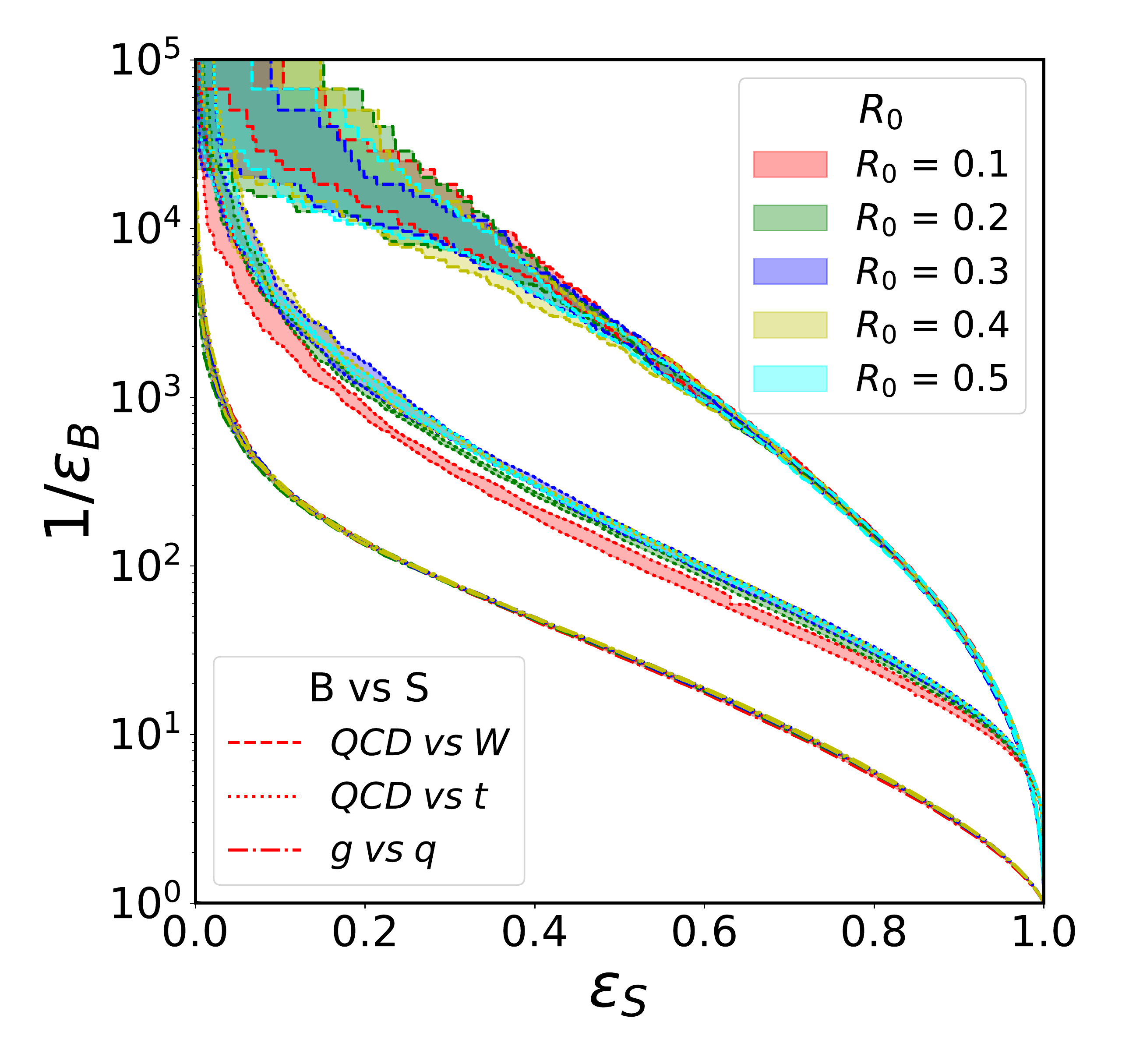}	
	\caption{ROC curve for the three tagging scenarioes and different values of $R_0$. The three sets of curves correspond to QCD vs $W$, QCD vs top, and gluons vs quarks from respectively from top to bottom. The band shows the maximum and the minimum values of the inverse of background acceptance $1/\epsilon_{B}$, for fixed values of signal efficiency $\epsilon_S$ from seperate runs. }		
	\label{fig:roc}
\end{figure*}
\begin{figure*}[t]
	\centering	
	
	\includegraphics[scale=0.285]{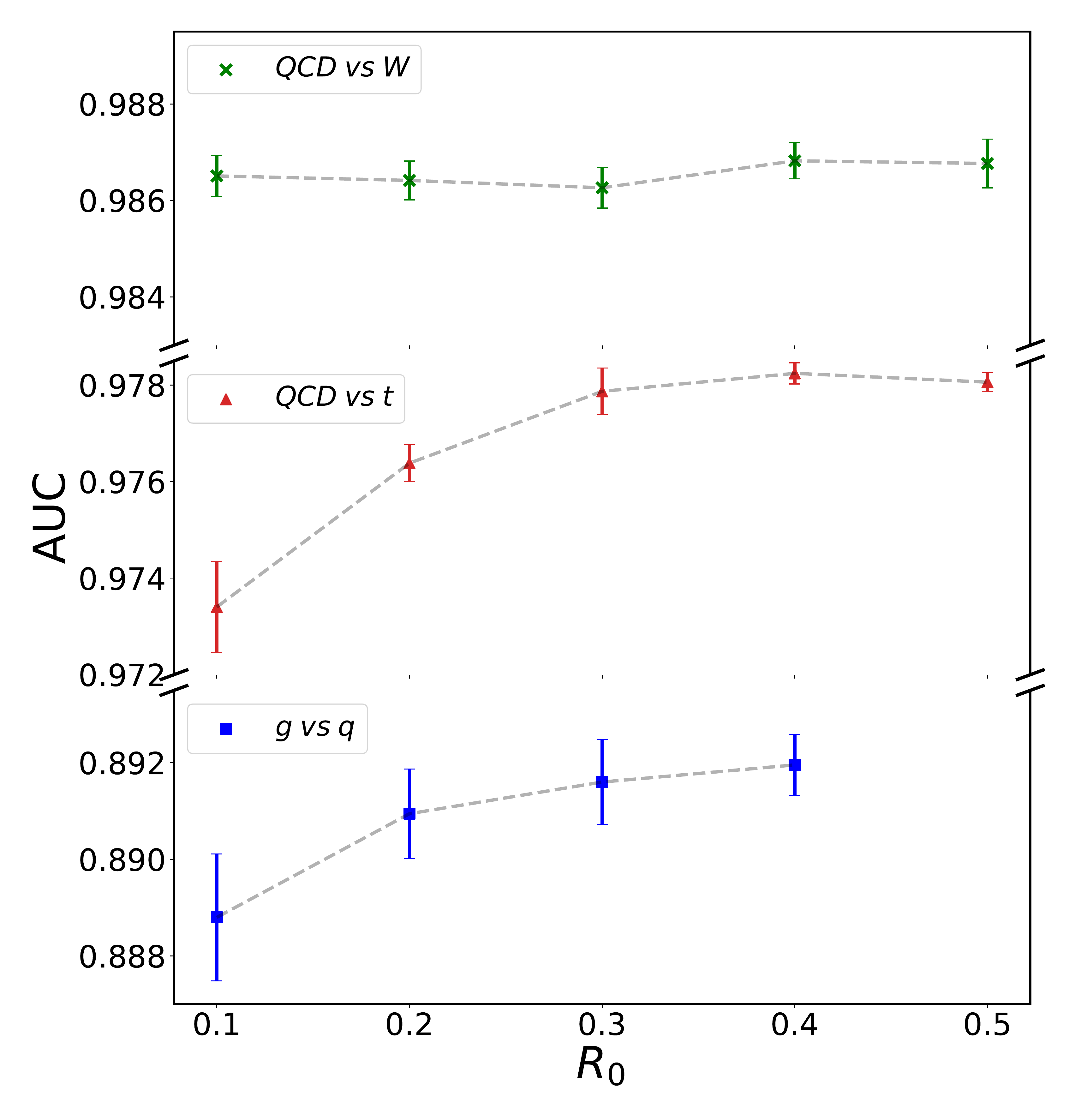}
	\caption{Variation of mean AUC with $R_0$ for the three tagging scenarios. For the $W$ tagging scenario (green cross), the AUC has saturated at $R_0=0.1$ and does not increase when compared to the other two. AUCs for the top vs QCD (red triangle), and the gluon vs quark (blue square) classification increases as we increase $R_0$. The error bands are the standard deviation from five training instances. }		
	\label{fig:auc}
\end{figure*}

The ROC curve for the three jet-tagging scenarios for the various values of $R_0$ are shown in Figure~\ref{fig:roc}. We evaluate the background acceptance $\epsilon_B$, at the same set of signal efficiencies $\epsilon_S$ for all instances of the trained networks. For a specific tagging case and fixed $R_0$, the boundary indicates the maximum and the minimum values of $1/\epsilon_B$ from the five training instances. The variation of the mean AUC and their error for the five training instances for each $R_0$ and the three cases are shown in Figure~\ref{fig:auc}. These values, along with the background rejection $1/\epsilon_B$ at 50\% signal efficiency, are shown in Tables \ref{tab:auc_val_qg}, \ref{tab:auc_val_top} and \ref{tab:auc_val_w} for gluon/quark, top and $W$ tagging respectively. For comparison, we also include relevant numbers for gluon vs quark and top tagging scenarios from Ref.~\cite{Komiske:2018cqr} for Energy Flow Networks (EFNs). Since we have not preprocessed our data, the values for top discrimination is for the unprocessed case.  
\begin{table}[t]
	\begin{center}
		\begin{tabular}{|c|c|c|c|}
			\hline
			Sl.No&$R_0$&AUC& $1/\epsilon_B$ at $\epsilon_S=50$\% \\
			\hline
			\multicolumn{4}{|l|}{$L=1$}\\
			\hline 
			1. &(EFN~\cite{Komiske:2018cqr})&0.8824$\pm$0.0005&28.6$\pm$0.3\\
			2. &0.1&0.8888$\pm$0.0013&30.1$\pm$0.3\\
			3. &0.2&0.8909$\pm$0.0009&30.1$\pm0.2$\\
			4. &0.3&0.8916$\pm$0.0008&30.7$\pm0.2$\\
			5. &0.4&0.8919$\pm$0.0006&31.0$\pm0.1$\\
			\hline
			\multicolumn{4}{|l|}{$L=2$ (Discussed Later)}\\
			\hline 
            1. &0.1&0.8932$\pm$0.0006&30.8$\pm0.2$\\
			\hline 	
		\end{tabular} 
	\end{center} 
	\caption{AUC values and the background rejection for different values of $R_0$ for \emph{gluons vs quark} tagging dataset. Uncertainties for AUC are the standard deviation from five training instances, while for the background rejection $1/\epsilon_B$ are half of the inter-quartile range. The first entry is quoted from the cited reference. }
	\label{tab:auc_val_qg} 
\end{table} 
The quark-gluon tagging networks already show improvement at $R_0=0.1$ with an AUC of 0.8888 over EFNs with 0.8824. However, for the top tagging case, the AUC (0.9734) for $R_0=0.1$ is less then that of EFNs (0.9760). This decrease indicates that the local structural information at that scale does not help distinguish QCD jets from top jets with a single message-passing operation. The local information learned by the message-passing phase confuses the downstream MLP, decreasing its performance compared to EFNs. Although, the message function or the downstream MLP we used is not exactly the same as the analogous per-particle map and the downstream MLP used in Ref.~\cite{Komiske:2018cqr}, and hence the comparison is not exactly like-for-like. The difference reaches parity at $R_0=0.2$, which further increases and reaches a stable value for higher $R_0$. Thus, for both scenarios, the energy-weighted message-passing help in better feature extraction of the local features. For the $W$ tagging results, we see very stable values of AUC (see Figure~\ref{fig:auc}), which do not vary appreciably with $R_0$ compared to the other two cases. The EMPN can already extract very rich features for the graphs at $R_0=0.1$, giving an AUC of 0.9865.  Increasing the complexity of the graph by enlarging $R_0$ does not add new information which the current architecture can extract. The stability of the AUC shown in Figure \ref{fig:auc} is likely due to the high kinematic range of the jets compared to the $W$ mass, giving the separation between the two decay products as $\Delta R\sim 2m_W/p_T\sim0.25$. To check whether the performance decreases for smaller $R_0$, we repeat the training process for $R_0=0.02$ and find that the mean AUC indeed falls mildly to 0.9845 for five training instances.
\begin{table}[t]
	\begin{center}
		\begin{tabular}{|c|c|c|c|}
			\hline
			Sl.No&$R_0$&AUC& $1/\epsilon_B$ at $\epsilon_S=50\%$\\
			\hline 
						1. &(EFN~\cite{Komiske:2018cqr})&0.9760$\pm$0.0001&143$\pm$ 2\\
			2. &0.1&0.9734$\pm$0.0009&115$\pm2$\\
			3. &0.2&0.9764$\pm$0.0004&151$\pm2$\\
			4. &0.3&0.9779$\pm$0.0005&167$\pm4$\\
			5. &0.4&0.9782$\pm$0.0002&174$\pm2$\\
			6. &0.5&0.9781$\pm$0.0002&168$\pm3$\\
			\hline 	
		\end{tabular} 
	\end{center} 
	\caption{AUC values and the background rejection for different values of $R_0$ for \emph{top tagging} dataset. Uncertainties for AUC are the standard deviation from five training instances, while for the background rejection $1/\epsilon_B$ are half of the inter-quartile range. The first entry is quoted from the cited reference.}
	\label{tab:auc_val_top} 
\end{table}

Other than the apparent variation of the mean AUCs and the ROC curves, we also see interesting features in the error bars of the AUC and the band of the ROC curves. If the AUC increases, its errors also gradually decrease as one increases $R_0$. On the other hand, across the different scenarios, the errors do not follow the same relation. The variation of AUC for each $R_0$ is due to the random initialisation of weights from the same underlying weight space with the same distribution\footnote{We are using the same initialiser for all networks.} for all networks. The optimisation proceeds via a gradient descent algorithm that goes to a local minimum of the loss function accessible from the initialised point in the space of weights for each instance. We can infer the relative quality of the local minima accessible from the initialised point. Lower the error, the easier it is to get to approximately similar values of the stable loss function. Comparing the three scenarios with stable AUC for the same $R_0$, we see that the top tagging case has a minor variance, followed by $W$ tagging. Thus, even though the performance of $W$ tagging is relatively higher, the distinguishing features for QCD vs top jets have a higher number of equally good local minima. The ROC band also enlarges with increased performance due to the decreasing statistics of the finite test sample.
\begin{table}[t]
	\begin{center}
		\begin{tabular}{|c|c|c|c|}
			\hline
			Sl.No&$R_0$&AUC& $1/\epsilon_B$ at $\epsilon_S=50$\% \\
			\hline
			1. &0.1&0.9865$\pm$0.0004&2415$\pm104$\\
			2. &0.2&0.9864$\pm$0.0004&2332$\pm95$\\
			3. &0.3&0.9863$\pm$0.0004&2381$\pm71$\\
			4. &0.4&0.9868$\pm$0.0004&2254$\pm80$\\
			5. &0.5&0.9868$\pm$0.0005&2300$\pm226$\\
			\hline 	
		\end{tabular} 
	\end{center} 
	\caption{AUC values and the background rejection for different values of $R_0$ for \emph{$W$ tagging} dataset. Uncertainties for AUC are the standard deviation from five training instances, while for the background rejection $1/\epsilon_B$ are half of the inter-quartile range.}
	\label{tab:auc_val_w} 
\end{table} 

\subsection{Examining IRC safety}
We now check the numerical stability of the network output for additional emissions. Since the network respects IRC safety, a jet with an additional splitting in the exact collinear or soft limit would have the same output without any splitting.  We explicitly verified that the difference between the network output of jets and their respective copies with one additional splitting in the exact collinear or soft limit are zero within numerical precision. In order to check the network output's stability, we create copies of an original top jet belonging to the test dataset by splitting the hardest constituent. We choose the hardest constituent since numerically, it should have the maximum effect on the probabilistic output due to the $\omega^{(\mathcal{K})}_i$ weighted node and graph readouts. The splitting is done as follows. We create a scaled copy $z_r\;p_q$ of the hardest four-momentum $p_q$. Taking the plane formed by the hardest particle and the softest particle in the jet, we rotate it by an angle $\theta$ giving us the four-momentum of one daughter $p_r$. The second daughter's four-momentum $p_s$ is determined by the enforcing conservation of energy and momentum. We vary the two quantities $z_r$ and $\theta$ independently to get the network output of the jet with an additional split $y_{\mathcal{S}'}(z_r,\Delta R_{rs})$ as a function of $z_r$ and $\Delta R_{rs}$.
\begin{figure*}[t]
	\centering	
	
	\includegraphics[scale=0.307]{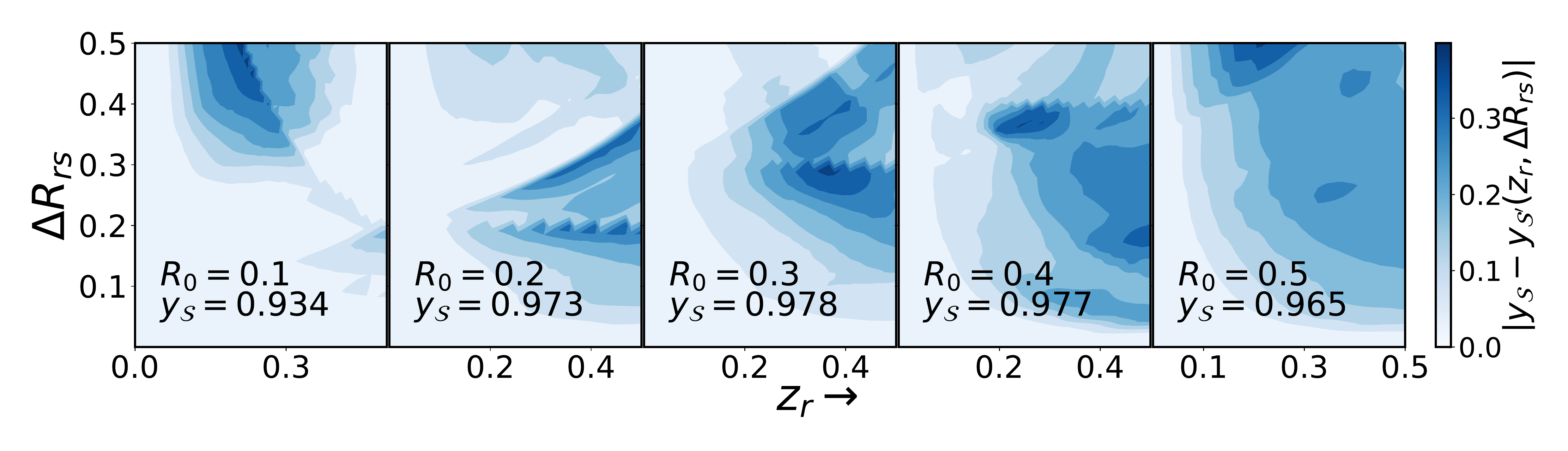}
	\caption{Variation of the network output with one additional particle, emitted from the hardest constituent in a top jet, on trained networks for various values of $R_0$. The contour figures show the contour of $|y_{\mathcal{S}}-y_{\mathcal{S}'}(z_r,\Delta R_{rs})|$ in the two-dimensional place $(z_r,\Delta R_{rs})$. Although, the differences are finite for non zero $z_r$ or $\Delta R_{rs}$, it goes to zero independently at the infra-red or collinear limits. }		
	\label{fig:irc_top_r}
\end{figure*}

The contour of the absolute difference $|y_{\mathcal{S}}-y_{\mathcal{S}'}(z_r,\Delta R_{rs})|$ between the network output of the initial jet $y_{\mathcal{S}}$ and those with an additional splitting $y_{\mathcal{S}'}(z_r,\Delta R_{rs})$ for different values of $R_0$ is shown in Figure~\ref{fig:irc_top_r}. We evaluate the difference of the best network from each of the five instances of training. For each $R_0$, we have plotted the contour having the maximum variance. The value of $y_{\mathcal{S}}$, which is the probability of the jet being a top, is also displayed. It can be seen that the difference goes to zero independently in the soft or collinear limits for all networks. This difference is low in considerable portions of the domain, indicating that the network output is relatively stable (at least for the particular jet).\footnote{We tested with multiple jets from the different classes and found similar features, for brevity we have only included a single plot.}  

We see an increase of the area with non-zero differences as one increases $R_0$. To understand this behaviour, we examine how the neighbourhood sets of each particle in the jet with an additional splitting evolve as one increases $R_0$. For a fixed $\Delta R_{rs}$, the two daughter's neighbourhood set would grow as one increases $R_0$. In contrast, for the remaining particles, the number of particles that have either of the daughter particles in them would increase with increasing $R_0$. Since they are greater in number, we expect this second aspect to influence the network output to a greater degree. Thus, even though the network performances are generally lower for smaller values of $R_0$, the network output is more stable for additional emissions that are not too soft or collinear. \emph{Increasing $R_0$, therefore, increases network performance at the cost of increased computational load (due to the addition of edges) but decreases the network's stability to QCD emissions.}

 Since the increase in performance for increasing $R_0$ comes at a price of a growing sensitivity to additional emissions, it is worth investigating how a deeper EMPN with more message-passing operations (which should increase the discrimination for a fixed $R_0$) fare against the same QCD radiations. We, therefore, train EMPN with two different message passing operations for $R_0=0.1$. For demonstration, we chose the gluon vs quark scenario because both classes' one-prong nature elevates the importance of the differing soft radiation patterns. We keep the structure of the downstream MLP and the message function at the first layer identical to the previously presented network. The second one is chosen to correspond to an edge-convolution operation given in Eq.~\ref{eq:edge_conv}, with $l=1$ in the superscript instead of $l=0$. The MLP, therefore, takes a 256-dimensional input and outputs a 128-dimensional vector. It contains two hidden layers of 128 nodes each, with {\tt ReLu} activation. The training is done five times with the same set of hyperparameters. We find a mean AUC of $0.8932\pm0.0006$ over the five training instances, confirming our presumption implying increasing performance with deeper models. Moreover, from Table~\ref{tab:auc_val_qg}, one finds that the value is even better than $R_0=0.4$ at $L=1$ with AUC=0.8919$\pm0.0006$, which indicates that the performance scales much faster with the number of message-passing operations $L$ than with $R_0$. 
	
\begin{figure*}[t]
	\centering	
	\includegraphics[scale=0.25]{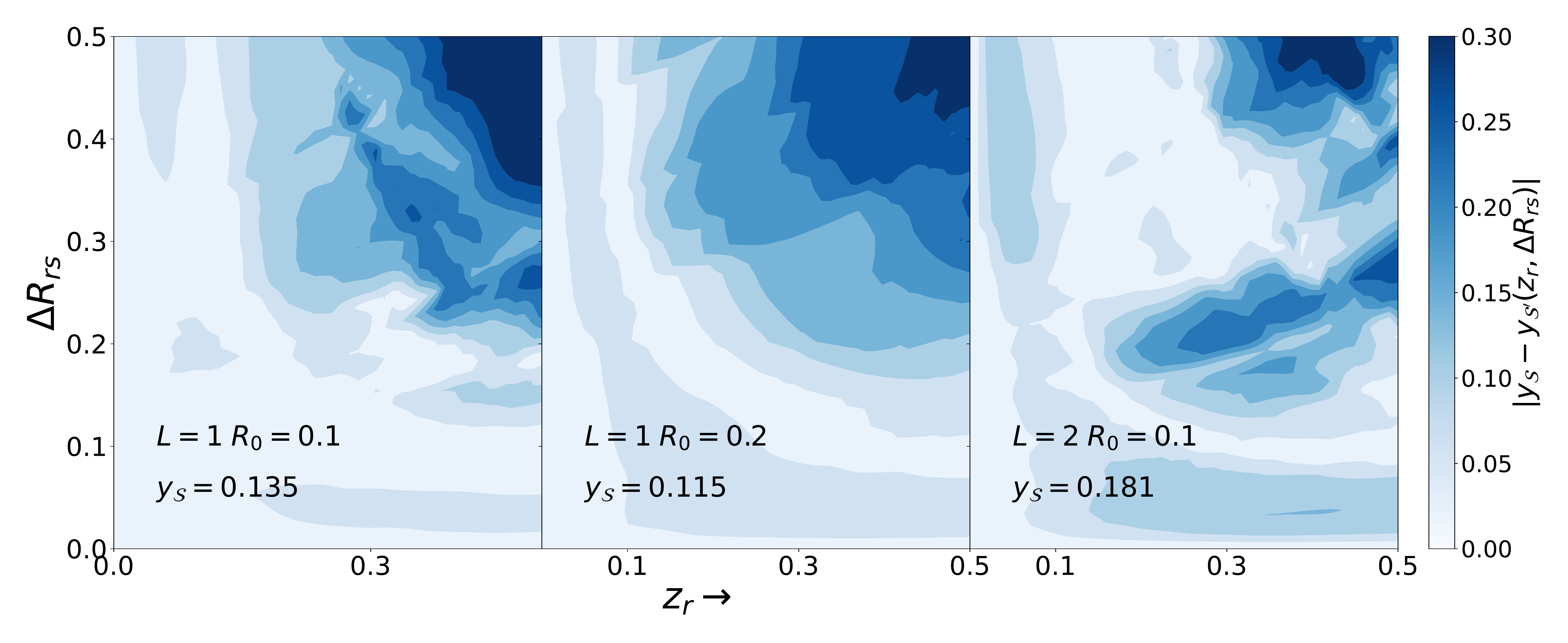}
	\caption{Comparing the variation of the network output $|y_{\mathcal{S}}-y_{\mathcal{S}'}(z_r,\Delta R_{rs})|$ of a gluon jet  for a deeper network (right) with $L=2$ and smaller radius $R_0=0.1$ against a shallower network (middle) with $L=1$ and different $R_0\in\{0.1,0.2,0.4\}$. For comparison, the variation with $L=1$ and $R_0=0.1$ is also shown on the left.}		
	\label{fig:irc_l1l2}
\end{figure*}

	We now turn to investigate the phenomenologically important resilience to additional emissions. Following the procedure explained in the preceding paragraphs, the contours of $|y_{\mathcal{S}}-y_{\mathcal{S}'}(z_r,\Delta R_{rs})|$ for a gluon jet with $(L,R_0)\in\{(1,0.1),(1,0.2),(2,0.1)\}$ are shown in Figure~\ref{fig:irc_l1l2}. Along with the increasing discrimination, the model with $L=2,\; R_0=0.1$ also provides better stability to additional emissions. The variation reduces with increasing depth for a constant $R_0$. Naturally, compared to $R_0=0.2$, which is less stable than with $R_0=0.1$ for constant $L$, we find that increasing $L$ has an overall better phenomenological suitability than increasing $R_0$. Thus, deeper networks increase the performance and enhance the stability of additional emissions. This stability might be due to the larger number of functional compositions that a deeper model applies to the input, thereby reducing the sensitivity of the weight space (fixed after the training) to perturbations in the data. However, it needs a more detailed study since our present analysis is not extensive and does not reflect a truly realistic QCD picture.

\section{Summary and Conclusions}
\label{sec:conclusion}
In hadron collisions, observables have to be defined in an infrared-safe way not to lose theoretical control. For example, the mass of a jet is defined through an elaborate jet reconstruction algorithm and can vary depending on the choice of such an algorithm. If the definition of perturbatively calculated observables is infrared-unsafe, a reliable comparison of the theoretically computed observable with an experimental measurement is at stake. 

With the progression towards highly performant yet increasingly elaborate, machine-learning-based reconstruction techniques, in recent years, a growing number of collider searches have relied on outputs of neural networks to perform classification tasks, i.e. to categorise if an event looks more like a signal or more like background. However, while such novel reconstruction techniques increase the sensitivity in searches for new physics, the experimental results depend crucially on a detailed understanding of their uncertainty budget. 

In strong similarity to jet observables, e.g., jet mass, the neural network output must be defined through an infrared-safe network algorithm. Concretely, if the neural network output changes when a soft or collinear splitting modifies the hadronic final state, the neural network algorithm is IRC unsafe. Thus, the neural network output becomes incalculable with perturbative techniques and is also sensitive to soft or collinear splittings in the parton shower and the choice of hadronisation model, which are often plagued by large uncertainties in the simulated training data. 

Thus, we propose an inherently infrared-safe definition for an Energy-weighted Message-Passing Network to mitigate becoming sensitive to IR uncertainties while maintaining a high classification efficacy. By defining local energy weight factors for the messages at the node readout, and global weight factors for the node features at the graph readout, we ensure that any generic message-passing function which takes the directional coordinates as inputs results in an IRC-safe network output when applied to graphs that remain structurally invariant in the presence of soft or collinear particles. Moreover, the operations need not be restricted to a single instance as repeating the weighted message-passing any number of times on the updated node features still guarantees IRC safety of the network output. 

Using an IRC-safe EMPN algorithm, we have applied this approach to the discrimination of hadronically decaying top quarks and W bosons from QCD jets and the classification of jets into quark or gluon-induced jets. We find this algorithm to be highly performant, at par with other state-of-the-art neural network classification methods quoted in the literature. Thus, our definition of an IRC-safe Energy-weighted Message-Passing Network paves the way to highly performant jet classification algorithms that are at the same time insensitive to often poorly modelled parts of the event simulation, i.e. phase-space regions in the training event samples that are plagued by large theoretical uncertainties.

\section*{Acknowledgements}
M.S. is supported by the STFC under grant ST/P001246/1. PK and VSN are supported by the Physical Research Laboratory (PRL), Department of Space, Government of India. Computational work was performed using the HPC resources (Vikram-100 HPC) and TDP project at PRL. PK and VSN thank Namit Mahajan and Satyajit Seth for useful discussions regarding the work. 

\appendix    
\bibliographystyle{JHEP}
\bibliography{ref}
	
\end{document}